\documentclass[showpacs, pra, twocolumn, a4paper]{revtex4-1}
\usepackage{amsmath}
\usepackage{amsfonts}
\usepackage{amssymb}
\usepackage{graphicx}
\usepackage{siunitx}
\usepackage{subfigure}
\usepackage{bm}
\usepackage{hyperref}
\usepackage{mhchem}
\usepackage{todonotes}
\usepackage{tikz,pgffor}
\usetikzlibrary{shadows}
\usetikzlibrary{calc}
\usetikzlibrary{%
    decorations.pathreplacing,%
    decorations.pathmorphing%
}
\graphicspath{{Figures/}}
\DeclareGraphicsExtensions{.pdf,.png} 

\newcommand{\ui}{\mathrm{i}}
\newcommand{\ue}{\mathrm{e}}
\newcommand{\spol}{\ensuremath{s}}
\newcommand{\ppol}{\ensuremath{p}}
\newcommand{\tvec}[1]{\ensuremath{\mathbf{#1}}}
\newcommand{\htvec}[1]{\ensuremath{\mathbf{\hat{#1}}}}
\newcommand{\kp}{\tvec{k}_\parallel}
\newcommand{\Enull}[1]{\ensuremath{\tvec{E}_0 \parallel \htvec{#1}}}

\begin{document}
\title{Plasmonic resonances at interfaces patterned by nanoparticle lattices}

\author{P. A. Letnes}
\email[]{paul.anton.letnes@gmail.com}
\author{I. Simonsen}
\email[]{ingve.simonsen@ntnu.no}
\affiliation{Department of Physics, The Norwegian University of Science and Technology (NTNU), NO-7491 Trondheim, Norway}
\author{D. L. Mills}
\email[]{(Deceased)}
\affiliation{Department of Physics and Astronomy, University of California, Irvine, California 92697, USA}

\date{\today}

\begin{abstract} 
We present theoretical studies of the nature of the collective plasmon resonances of surfaces upon which ordered lattices of spherical metallic particles have been deposited. The collective plasmon modes, excited by light incident on the surface, are explored for both square and rectangular lattices of particles. The particular resonances excited by an incident beam of light depend on the frequency, polarization, and angles of incidence. We show that one can create surfaces for which the polarization of the reflected light is frequency dependent. The form of the polarization dependent spectra can be tuned by choosing materials and the parameters of the nanoparticle array.
\end{abstract} 


\maketitle

\section{\label{sec:Introduction} Introduction} 

It is of great interest to create surfaces whose response to incident light can be controlled to produce desirable effects of various sorts. This may be accomplished in diverse ways. For instance one may modulate the surface profile. Very striking and remarkable reflectivity properties are found in systems containing surfaces whose roughness is prepared according to prescriptions provided by theoretical simulations~\cite{maradudin2008designer}. We refer the reader to a recent review wherein the scattering of light from roughened surfaces is discussed in detail~\cite{springerlink:10.1140/epjst/e2010-01221-4}.

In addition to modulating the profile of a surface in a controlled fashion, one may also deposit material onto it in a structured manner. For instance, to enhance photocurrents in solar energy devices, metallic nanoparticles are deposited on or near the surface of solar devices~\cite{Pillai:2007dn,Atwater:2010uq}. We remark that the nanoparticle arrays in these systems are very highly disordered. On a related note, disordered structures generated by laser bombardment has been shown to modify the color of metallic surfaces in a profound way~\cite{vorobyev:041914}.

Recently, progress has been made towards using the plasmonic properties of nanowires~\cite{doi:10.1021/nl0808435,Ghadyani:11} or nanoparticles~\cite{moiseev:79931Q,aas:80822W} as optical polarizers, in a fashion reminiscent of wire-grid polarizers. This work promises to make polarizers that are microscopic in size, and which possess other interesting properties. For instance, the spectral response of nanoparticles can be tuned according to their shape~\cite{doi:10.1080/01442350050034180} or internal structure~\cite{2011arXiv1112.1126M}, allowing the fabrication of optical components with new and interesting properties. In all these studies, the properties of structures on the nanoscale, notably nanoparticles, lead to macroscopic optical effects~\cite{Kreibig95,hulst1981,Bohren:2007fk}.

In this paper, we explore a very different kind of nanoparticle-coated surface, namely, a dielectric substrate upon which an ordered array of sub-wavelength metallic nanoparticles~\cite{LPOR:LPOR200810003} has been deposited~\cite{0957-4484-19-41-415702,doi:10.1021/la7016209,liu:2030}. As we shall see from the calculations presented below, such a surface exhibits striking optical properties that may be tuned by varying the character of the nanoparticle array.

There are two important features of such an array. First, let $\ell$ be a length that characterizes the size of the unit cell, and assume that light of (vacuum) wavelength $\lambda$ illuminates the structure. If $\lambda / \ell > 1$, a condition satisfied for sub-wavelength arrays, there will only be a single reflected specular beam, very much as realized for a perfectly flat surface.
For particle arrays with linear dimensions in the range of the wavelength or larger ($\ell \gtrsim \lambda$), one realizes additional reflected beams in the form of Bragg waves. As the distance between the nanoparticles decreases into the sub-wavelength regime, the additional Bragg waves collapse into evanescent waves confined to the near vicinity of the surface, leaving a single specular reflected beam, once again very much as for a perfectly smooth surface.

A metallic nanoparticle array also supports plasmonic resonances that may be excited by incident light. For an isolated nanoparticle of radius $a \ll \lambda$ made from a ``plasmonically active'' material such as silver (\ce{Ag}), the optically active plasmonic (Mie) resonance lies in the ultraviolet (at $\hbar\omega \approx \SI{3.5}{\electronvolt}$ for \ce{Ag}, where $\omega$ is the angular frequency of the incident light). Depositing particles onto a substrate will cause inter-nanoparticle and particle-substrate interactions that redshifts the Mie resonance so it can lie in the visible, as illustrated by previous work; see Fig.~9 in Refs.~\onlinecite{PhysRevB.83.075426,PhysRevB.85.149901}.

More important for the present study is the role of interactions between nanoparticles in a dense array. These produce collective plasmonic bands whose dispersion relations and effects on polarization are controlled by the properties of the array. As we shall see, these collective plasmon modes can be excited by the incident light, with the consequence that the reflectivity of the surface becomes highly dependent on both the frequency, the angles of incidence, and the polarization of the incident light. If the surface is illuminated with unpolarized light, the reflected light will be polarized, but the degree of polarization can be strongly frequency dependent, as illustrated by the calculations presented below. By altering the microstructure of the array, one may in principle tune the polarization characteristics of the reflected light. The purpose of this study is to explore these effects, and to demonstrate how such effects can be simulated in a way that is not computationally demanding.

A classic and frequently used model to obtain the optical response of nanoparticle patterned surfaces is the polarizable dipole model~\cite{Yamaguchi1974173}. In this model, each nanoparticle is treated as a polarizable point dipole, and the effect of the substrate is accounted for via image dipoles.   For nanospheres with intersphere spacing as small as those of our geometries, such an approximation is very poor. There are highly localized patches of induced charge localized around the points of closest contact between neighboring spheres, and also around the closest point of contact between the sphere and the substrate. One can appreciate this from our previous calculation of the spatial dependence of enhanced fields near an \ce{Ag} dimer on a dielectric surface, as illustrated in Fig.~10 of Refs.~\onlinecite{PhysRevB.83.075426,PhysRevB.85.149901}.
Also in Fig. 5 of Ref.~\onlinecite{Chu:2008gd} one sees a detailed presentation of the electric field near the point of closest contact between two spheres when they are excited at their plasmon resonance. It is essential to include higher order multipole moments in the description of interactions between nearby nanospheres, and between nanospheres and a substrate upon which they are deposited, as we have done in the results displayed below.

A more direct approach is that of the discrete dipole approximation (DDA)~\cite{1973pp}, which can be expanded to include retardation and substrate effects~\cite{Schmehl:97}. The DDA is a direct numerical method, relying on discretizing the volume of particles into polarizable dipoles. In contrast, the method used in this work relies on multipole expansion of the quasistatic potential~\cite{Jackson:1962aa}. The advantage of this method is that it gives results closer to analytical mathematics, as well as being computationally efficient, assuming that one does not need to include very high multipole orders.

The organization of this paper is as follows. Section~\ref{sec:theory} contains the necessary theoretical formalism. In Sec.~\ref{sec:collective-plasmon-modes}, we present our studies of the plasmon collective modes of arrays of \ce{Ag} particles supported by an aluminium oxide (alumina, \ce{Al2O3}) substrate, and in Sec.~\ref{sec:reflectivity} we discuss the optical reflectivity of two model systems. Concluding remarks are presented  in Sec.~\ref{sec:Conclusion}.


\section{\label{sec:theory}Theory} 

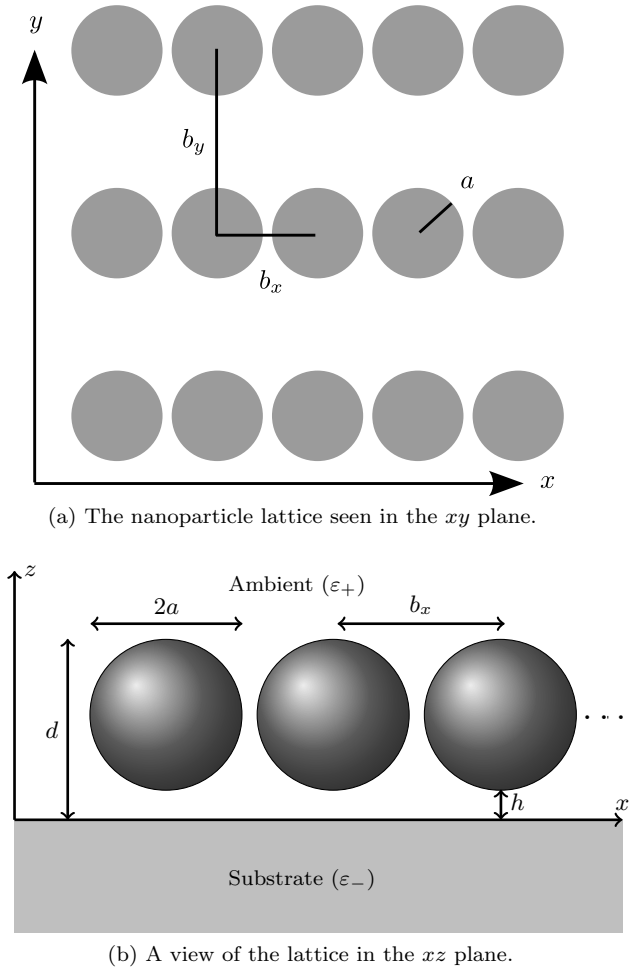
\begin{figure}[tb]
    \centering
    \subfigure[~The nanoparticle lattice seen in the $xy$ plane.]
    {\label{fig:geometry}
        \includegraphics{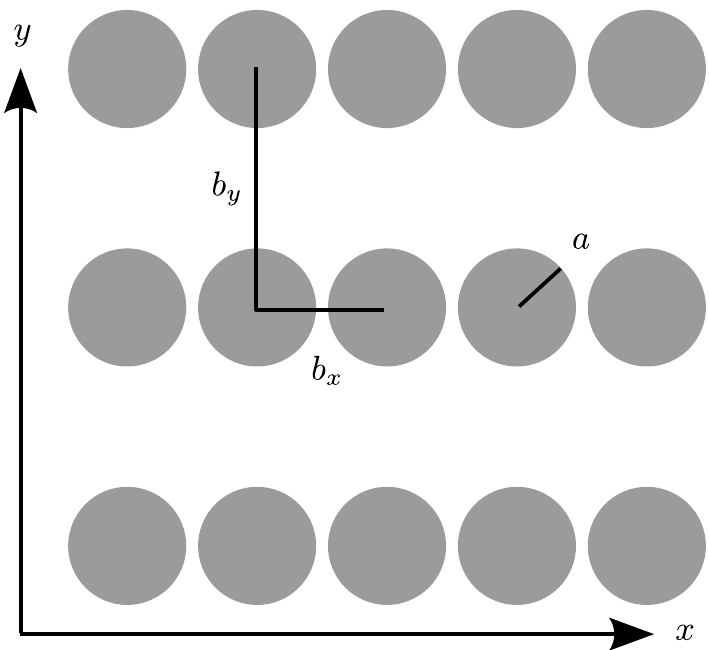}
    }
    \subfigure[~A view of the lattice in the $xz$ plane.]
    {\label{fig:geometry_2}
        \newcommand{\drawcircle}[3]{%
   \pgfmathparse{1/(2*#2*#2)}
   \let\tempradius\pgfmathresult
   \draw[fill=#3] (#1/#2,\tempradius) circle (\tempradius);
}
\def\fordN{5}
\def\fordA{0}
\def\fordB{1}
\def\fordC{1}
\let\fordD\fordN

\begin{tikzpicture}[
    media/.style={font={\footnotesize}},
    ]
    \fill[gray!50] (-4,-1.5) rectangle (4,0);
    \draw[->,black,line width=1.0pt](-4,0)--(4,0)
        node[above]{$x$};
    \draw[->,black,line width=1.0pt](-4,0.0)--(-4,3.3)
        node[right]{$z$};
    \path[right,media] (-1.3,3.1)  node {Ambient $(\varepsilon_+)$}
                 (-1.3,-0.8) node {Substrate $(\varepsilon_-)$};
    \foreach \x in {-2.2, 0.0, 2.2}
        \shadedraw [ball color = gray] (\x,1.2,-0.5) circle (1.0cm);
    \draw[<->,black,line width=1.0pt]
        (-3.0,2.6)--(-2.0,2.6)node[above]{$2 a$}--(-1.0,2.6);
    \draw[<->,black,line width=1.0pt]
        (0.25,2.6)--(1.35,2.6)node[above]{$b_x$}--(2.45,2.6);
    \draw[<->,black,line width=1.0pt]
        (2.40,0.0)--(2.40,0.24)node[right]{$h$}--(2.40,0.4);
    \draw[<->,black,line width=1.0pt]
        (-3.3,0.0)--(-3.3,1.2)node[left]{$d$}--(-3.3,2.4);
    \draw[-,black,line width=1.0pt]
        (3.8,1.36)--(3.8,1.36)node{\Large\ldots};
\end{tikzpicture}
    }
    \caption{\label{fig:both_geometry}
        (a) Nanospheres of radius $a$ are arranged on a rectangular lattice with lattice constants $b_x$ and $b_y$. The $z$ axis points out of the substrate, and we always assume $b_x \leq b_y$. (b) The ambient-substrate interface is located at $z = 0$. The definitions of $d$, the effective film thickness; $a$, the sphere radius; $b_x$, the lattice constant along \htvec{x}; and the parameter $h$ are indicated. The spheres are all characterized by the dielectric function $\varepsilon(\omega)$. Note that $h$ is exaggerated for clarity.
    }
\end{figure}

The system we study consists of a periodic array of non-overlapping \ce{Ag} nanoparticles supported by an \ce{Al2O3} substrate as depicted in Fig.~\ref{fig:both_geometry}. The global Cartesian coordinate system $\tvec{r} = (x,y,z)$ is chosen such that the plane $z = 0$ coincides with the (flat) surface of the substrate that is located in the region $z < 0$ and characterized by the dielectric function $\varepsilon_-(\omega)$. The ambient ($z > 0$) is assumed to be vacuum, and therefore $\varepsilon_+\equiv 1$.

A set of identical \ce{Ag} nanospheres of radius $a$ are arranged on a regular lattice close to the surface of the substrate.
For each spherical nanoparticle, we associate a position vector $\tvec{R}_{ij} = \left( x_{ij}, y_{ij}, h + a \right)$ pointing from the origin of the (global) coordinate system to the center of each particle, where the particles are indexed $i, j = 0, \pm 1, \pm 2, \ldots$. For later convenience, we assume a small but finite positive value for the parameter $h$~[Fig.~\ref{fig:geometry_2}].
The particle $i =j= 0$ is assumed to be located on the $z$-axis so that $\tvec{R}_{00} = (0, 0, a + h)$.
Furthermore, each sphere $ij$ has associated with it a (local) coordinate system $\mathcal{S}_{ij}$ that has its origin located at the center of that sphere and its axes oriented parallel to those of the main coordinate system. The position vector in $\mathcal{S}_{ij}$ we denote by $\tvec{r}_{ij} = \left( r_{ij}, \theta_{ij}, \phi_{ij} \right)$.

The \ce{Ag} nanospheres are characterized by the dielectric function $\varepsilon(\omega)$. Corrections to the dielectric function of the particles due to, e.g., finite size and temperature effects are necessary in order to obtain good agreement between theoretical predictions and experimental measurements~\cite{Simonsen:2003aa}. This will not be done here, however, since comparison to experimental data will not be our main concern. Hence, for reasons of simplicity, we have assumed bulk values for all dielectric functions.

In the following we shall consider both square and rectangular lattice structures. Without loss of generality, the coordinate system can be oriented so that the lattice constants $b_x$ and $b_y$, corresponding to the directions \htvec{x} and \htvec{y}, respectively, obey $b_x \leq b_y$ [see Fig.~\ref{fig:geometry}]. Here, a caret above a vector indicates that it is a unit vector.  Furthermore, since the spheres are non-overlapping, we also have that $2a < b_x$.

Although in general square lattices are formally a subset of rectangular lattices, we will here restrict the term ``rectangular lattice'' to lattices for which $b_x < b_y$. A comparison between square and rectangular lattices will provide us with an assessment of the range of electrostatic coupling between the spheres in the lattice. Moreover, in the limit $b_y \gg a$, where interactions between the spheres in the $\htvec{y}$ direction safely can be neglected, the system essentially consists of non-interacting, parallel, linear chains of nanoparticles~\cite{PhysRevB.68.245420}. Moreover, the word ``chains'' will be used about the lines of spheres parallel to the $x$ axis, as $b_x < b_y$.

\subsection{Multipole expansion} 
\label{sub:Multipole expansion}

In this work we will focus on systems for which $b_y \ll \lambda$, where the \emph{quasistatic approximation} applies~\cite{Jackson:1962aa}. Then, if all materials are assumed to be non-magnetic, the electromagnetic properties of the system is fully described by the electrostatic potential, $\psi(\tvec{r})$, satisfying the Laplace equation:
\begin{align*}
    \tvec{\nabla}^2 \psi(\tvec{r}) = 0.
\end{align*}
Also, the appropriate boundary conditions~\cite{Jackson:1962aa} on the surfaces of each sphere and at the interface between the substrate and the ambient have to be fulfilled. By definition, the electric field can be calculated from $\tvec{E}(\tvec{r}) = -\tvec{\nabla} \psi(\tvec{r})$~\cite{Jackson:1962aa}. Finally, we assume that the incident optical radiation can be modeled as a spatially uniform electric field $\tvec{E}_0$ of angular frequency $\omega = 2\pi c/\lambda$. As we are employing the quasistatic approximation, $\omega$ only appears in the frequency dependence of the dielectric functions $\varepsilon_{\pm}(\omega)$ and $\varepsilon(\omega)$.

In what follows, we adapt the \emph{multipole expansion} formalism presented in detail in Ref.~\onlinecite{PhysRevB.83.075426}, only modifying it where necessary to take into account the symmetries and the infinite nature of the lattice. The structure of the lattice requires the potential to satisfy the Bloch--Floquet theorem~\cite{kittel1976introduction}. Let $\psi_{ij}(\tvec{r}_{ij})$ denote the scalar potential of the particle located at $\tvec{R}_{ij}$ (in the global coordinate system) and expressed in terms of its local coordinate system, $\tvec{r}_{ij}$. We assume that the potential from each nanoparticle, $\psi_{ij}$, is identical save for a phase factor, due to the phase of the incident electric field:
\begin{align}
\label{eq:potential-translation}
    \psi_{ij}(\tvec{r}_{ij}) =
        \psi_{00}(\tvec{r}_{ij})\ue^{\ui\kp\cdot \Delta\tvec{R}_{ij}}.
\end{align}
In writing Eq.~\eqref{eq:potential-translation}, we have introduced the lattice vector $\Delta\tvec{R}_{ij} \equiv \tvec{R}_{ij}-\tvec{R}_{00}=(ib_x,jb_y,0)$ describing the periodic array of spheres, and $\kp$ denotes the component of the wave vector $\tvec{k}$ of the incident electric field that is parallel to the $xy$ plane: $\kp = \htvec{x}k_x +\htvec{y}k_y$. Equation~\eqref{eq:potential-translation} signifies that once the scalar potential of sphere $i = j = 0$ is known, it is essentially known for all spheres of the lattice.
This is a consequence of the Bloch--Floquet theorem, and the form~\eqref{eq:potential-translation} is similar to the tight binding description of electron energy bands in solids. The Bloch phase factor is also assumed to be present in the potential of the corresponding image multipole. In passing, we note  that  Eq.~\eqref{eq:potential-translation} also predicts that the potential $\psi_{ij}(\tvec{r}_{ij})$ is invariant under a replacement of  $\kp$ by $\kp+\tvec{G}_{mn}$ where $\tvec{G}_{mn}=2\pi(m/b_x,n/b_y,0)$ denotes a reciprocal lattice vector (with $m$ and $n$ integers).
This invariance follows from the fact that a scalar product of a primitive lattice vectors from direct space, and one from reciprocal space, equals an integer multiple of $2\pi$~\cite{kittel1976introduction}. Hence, for the sake of the calculation of the potentials, it suffices to consider wave vectors $\kp$ in the first Brillouin zone. Moreover, since this work considers the limit $\lambda\gg b_y\geq b_x$, one may take the limit $|\kp| \to 0$ when calculating the potentials.

In the formalism presented in Ref.~\onlinecite{PhysRevB.83.075426}, the total potential in the ambient, $\psi_+(\tvec{r})$, is expressed as a superposition of the potential corresponding to the background electric field ($\tvec{E}_0$), the potential scattered from each nanoparticle, and the potential from an image multipole designed to take the substrate into account:
\begin{align*}
    \begin{aligned}
        \psi_+ (\tvec{r}) = -\tvec{r} \cdot \tvec{E}_0
            &+ \sum_{i = -\infty}^{\infty}
                \sum_{j = -\infty}^{\infty}
                    \psi_{ij}(\tvec{r}_{ij}) \\
            &+ \sum_{\bar{i} = -\infty}^{\infty}
                \sum_{\bar{j} = -\infty}^{\infty}
                    \psi_{\bar{i}\bar{j}} (\tvec{r}_{\bar{i}\bar{j}}).
    \end{aligned}
\end{align*}
Here the indices $\bar{i}$ and $\bar{j}$ indicate that the quantity $\psi_{\bar{i}\bar{j}}$ represents contribution to the potential from image multipoles associated with the sphere $ij$, and $\tvec{r}$ is the position vector in the global coordinate system. With the shorthand notation $\sum_{lm} = \sum_{l = 0}^\infty \sum_{m = -l}^l$, the multipole expansion we use for the scalar potential is given by~\cite{Jackson:1962aa}:
\begin{subequations}
\label{eq:psi_j}
\begin{align*}
    \psi_{00} (\tvec{r}_{00})
        &=
            \sum_{lm} A_{lm} r_{00}^{-l-1}
                Y_l^m (\theta_{00}, \phi_{00}), \\
    \psi_{\bar{0}\bar{0}} (\tvec{r}_{\bar{0}\bar{0}}) &=
        \sum_{lm} A^{(R)}_{lm}
            r_{\bar{0}\bar{0}}^{-l-1}
            Y_l^m (\theta_{\bar{0}\bar{0}}, \phi_{\bar{0}\bar{0}}).
\end{align*}
\end{subequations}
Here, the symbols $A_{lm}$ and $A^{(R)}_{lm}$ are multipole expansion coefficients to be determined. The symbol $Y_l^m$ refers to the spherical harmonic functions as described by Ref.~\onlinecite{Jackson:1962aa}.
By symmetry arguments and by matching the boundary conditions at $z = 0$~\cite{bedeaux_book}, the coefficients $A_{lm}$ are related to the coefficients $A_{lm}^{(R)}$ by the relation
\begin{align*}
    A_{lm}^{(R)} &=
        (-1)^{l + m}
        \frac
            {\varepsilon_+ - \varepsilon_-}
            {\varepsilon_+ + \varepsilon_-}
        A_{lm}.
\end{align*}
We note that the sum over $l$ for practical reasons has to be truncated at a finite value $L_{\mathrm{max}}$, meaning that only terms for which $l \leq L_{\mathrm{max}}$ are included in the sum.
 It should be stressed that local fields may not have converged fully for this cutoff; however, the observed collective resonances and reflectivity are less sensitive to the details of the local fields. This is especially true when considering large arrays of metallic nanoparticles, in which the interparticle interactions are stronger than the particle-substrate interactions. The potential originating from other nanoparticles (and images), $\psi_{ij}$, can be found from $\psi_{00}$ and Eq.~\eqref{eq:potential-translation}. In order to determine the $A_{lm}$ coefficients, one forms a linear system of equations which couples all multipole orders $l$. How to form and solve this system of equations is discussed in detail in Ref.~\onlinecite{PhysRevB.83.075426}.

Although we are working with lattices that have discrete translational symmetry with respect to the lattice vector, there is little to be gained by use of a Fourier representation of the relevant lattice sums. In mathematical terms, the vector $\tvec{R}_{\bar{i}\bar{j}} - \tvec{R}_{kl}$ pointing from sphere $kl$ to image $\bar{i}\bar{j}$ does not lie in the $xy$ plane: Its polar angle will vary depending on the relative position of the image multipoles and the sphere on the $z$-axis. Hence, the presence of the substrate, and thus the image multipoles, leads to terms in the lattice sums lacking the symmetry needed to use the Fourier transform. For this reason, the necessary sums over spheres~\cite{PhysRevB.83.075426} are performed directly in real space, up to $N = 10$ unit cells away from the sphere on the $z$-axis. We stress that these sums include both nanoparticles and multipole images out to $N$ unit cells.

In all of our simulations, we have kept a finite value of the parameter $h$ to ensure convergence of the spherical harmonics expansion. This is because a singularity arises at the bottom of the sphere if we choose $h = 0$~\cite{Romero:2006rm}. In the simulations we present here, we have elevated the layer of spheres off the substrate by the amount $h = 0.01a$. We then find very good convergence in terms of the collective excitations and reflectivity behavior. It should be stressed that global behavior, such as the reflectivity of the lattice, does not depend on the details of $h$, even though the convergence of the local fields is more sensitive to the parameters $L_{\max}$ and $h$.

As a proxy for the optical response of nanoparticles or clusters of nanoparticles, we use the dipole moment, $\tvec{p}$, or its
dimensionless analog defined as (in SI units)~\cite{PhysRevB.83.075426}
\begin{subequations}
\label{eq:p-bar}
\begin{align}
    \tvec{\bar{p}}(\omega) =
        \frac{\tvec{p}(\omega)}{a^3 \varepsilon_0 E_0},
\end{align}
and its absolute value
\begin{align}
    \bar{p}\left( \omega \right)
    = \left(
        \tvec{\bar{p}}^\dagger \tvec{\bar{p}}
    \right)^{1 / 2},
\end{align}
\end{subequations}
where the superscript $\dagger$ on a quantity indicates its Hermitian conjugate. The components of the dipole moment vector can be found from~\cite{PhysRevB.83.075426}
\begin{subequations}
\label{eq:dipole-moment-components}
    \begin{align}
        \bar{p}_x &=
            \left( \frac{3}{8 \pi} \right)^{1/2}
                \frac{A_{1,-1} - A_{1,1}}{a^2}, \\
        \bar{p}_y &= -\mathrm{i}
            \left( \frac{3}{8 \pi} \right)^{1/2}
            \frac{A_{1,-1} + A_{1,1}}{a^2},
     \end{align}
and
     \begin{align}
        \bar{p}_z &=
            \left( \frac{3}{4 \pi} \right)^{1/2}
            \frac{A_{1,0}}{a^2}.
    \end{align}
\end{subequations}
The $i$th component of the dipole moment is related to the incident electric field ($\tvec{E}_0$)  in the vicinity of the particle by
\begin{align}
    \label{eq:polarizability-definition}
    p_i = \sum_{j = 1}^3 \alpha_{ij}E_{0,j},
\end{align}
where $\alpha_{ij}$ denotes the polarizability tensor of one of the  nanoparticles.

In passing we note that even though the dipole moment only depends on the lowest order ($l = 1$) expansion coefficients, it carries information on higher order resonances, as the system of equations for the $A_{lm}$ coefficients couple coefficients of all orders~\cite{PhysRevB.83.075426}. Moreover, for ease of comparison, we will below always refer to the dipole moment of a single particle.


\subsection{\label{sub:reflectivity} Surface reflectivity} 

The most readily observable quantity that gives an indication of plasmonic activity is the surface reflectivity. The reflectivity of the system we consider can be calculated via several routes. The first approach that will be mentioned is due to Bedeaux and Vlieger~\cite{bedeaux_book} (see also Refs.~\onlinecite{PhysRev_61_7722_2000,Lazzari_24_267_2001,PhysRevB_65_235424_2002,GranFilm,Simonsen:2003aa,Remi_Nanotech}). It introduces an equivalent geometry consisting of the same ambient and substrate as the original geometry, but without the nanoparticles. The influence of the latter is accounted for through effective boundary conditions for the electromagnetic field on the ambient--substrate interface. These boundary conditions depend on so-called surface susceptibilities that contain, for instance, the effect of the size, shape, aspect ratio, and location of the nanoparticles. The surface susceptibilities are obtained from the multipole coefficients $A_{lm}$ (Sec.~\ref{sub:Multipole expansion}). The approach of Bedeaux and Vlieger~\cite{bedeaux_book,PhysRev_61_7722_2000,PhysRevB_65_235424_2002} has proven to produce accurate results for the surface reflectivity that compares quantitatively well to experimental measurements~\cite{PhysRev_61_7722_2000,PhysRevB_65_235424_2002,Remi_Nanotech}. It has been used successfully to interpret and invert  experimental reflectivity data~\cite{Simonsen:2003aa,Remi_Nanotech}. Recently, this has opened for the possibility of detailed \emph{in situ} and real time studies of the growth of supported nanoparticles during deposition~\cite{Remi_Nanotech}. However, the accuracy of the Bedeaux--Vlieger method comes at a cost: The approach is somewhat technical, and may be challenging to grasp for those not familiar with it.

Since our main concern in this study is not quantitative interpretation of experimental reflectivity measurements, we will instead follow a simpler and more qualitative route towards obtaining the surface reflectivity. To this end, we define above the substrate a thin film region of thickness $d$~[Fig.~\ref{fig:geometry_2}] that contains the nanoparticles, and when it is homogenized~\cite{Book:Choy-1999,Book:Milton-2002,Bergman1978}, an effective medium results.
The classic Maxwell Garnett theory~\cite{garnett1904colours,azzam1987ellipsometry,Book:Choy-1999,Book:Milton-2002,Bergman1978} or the Bruggeman theory~\cite{Bruggeman1935} for such films are essentially only sensitive to the volume fraction of particles and the (bulk) dielectric functions of the materials of the particles and substrate, and not to the shape and environment (e.g., the ambient-substrate interface) surrounding the particles of the lattice. Instead of employing the Maxwell Garnett or Bruggeman theories directly, we construct an effective medium theory that does depend on other parameters, such as the local environment, via the calculated polarizability of the metallic particles.
The starting assumption of our effective medium model is that the lattice of nanospheres can be represented as an anisotropic thin film of thickness $d = 2a + h \approx 2a$~[Fig.~\ref{fig:geometry_2}].

Due to the symmetry properties of the two-dimensional rectangular (or square) lattice,
the dielectric tensor $(\stackrel{\leftrightarrow}{\varepsilon})$ will in our choice of coordinate system be diagonal:
\begin{align}
    \stackrel{\leftrightarrow}{\varepsilon} = \begin{pmatrix}
        \varepsilon_x & 0 & 0 \\
        0 & \varepsilon_y & 0 \\
        0 & 0 & \varepsilon_z
    \end{pmatrix}.
    \label{eq:dielectric_tensor}
\end{align}
where $\varepsilon_x$, $\varepsilon_y$, and $\varepsilon_z$ are the principal dielectric constants. Hence, the polarizability tensor, $\stackrel{\leftrightarrow}{\alpha}$, must also be diagonal. The macroscopic polarization, $\tvec{P}$, defined as the total dipole moment per unit volume, is thus given by
\begin{align}
    \label{eq:polarizability-per-volume}
    P_i = \frac{\varepsilon_0}{b_x b_y d} \sum_{j = 1}^3 \alpha_{ij} E_{0, j},
\end{align}
where $P_i$ denotes the $i$th component of $\tvec{P}$, $\alpha_{ij}$ denote the polarizability tensor components, and subscript indices $i = 1, 2, 3$ correspond to subscripts $x, y, z$, respectively. Moreover, $b_x b_y d$ is the volume of the effective thin film covering one unit cell of the 2D lattice. In SI units, the displacement field, \tvec{D}, is given by
\begin{align}
    \label{eq:d-definition}
    \tvec{D} = \varepsilon_0 \tvec{E} + \tvec{P}.
\end{align}
By inserting Eq.~\eqref{eq:polarizability-per-volume} into Eq.~\eqref{eq:d-definition}, we obtain
\begin{align*}
    \begin{aligned}
        D_i &= \varepsilon_0 E_{0, i} + P_i
            = \varepsilon_0 \left( E_{0, i} + \frac{1}{b_x b_y d}
                \sum_{j = 1}^3\alpha_{ij} E_{0, j} \right) \\
            &= \varepsilon_0 \sum_{j = 1}^3
            \left(
                \delta_{ij} + \frac{1}{b_x b_y d} \alpha_{ij}
            \right) E_{0, j}
            = \varepsilon_0 \sum_{j = 1}^3 \varepsilon_{ij} E_{0, j},
    \end{aligned}
\end{align*}
so that we get for the component of the dielectric tensor
\begin{align}
  \label{eq:dielectric_tensor_components}
    \varepsilon_{ij} = \delta_{ij} \
            + \frac{1}{b_x b_y d} \alpha_{ij}.
\end{align}
The polarizability tensor components, $\alpha_{ij}$, that appears in Eq.~\eqref{eq:dielectric_tensor_components}, can be related to the multipole coefficients $A_{lm}$ by comparing Eq.~\eqref{eq:polarizability-definition} to the multipole expansion of $\psi(\tvec{r})$~\cite{Jackson:1962aa}. This yields the formulas
\begin{subequations}
\label{eq:polarizability-from-alm}
    \begin{align}
        \label{eq:alpha_xx}
        \alpha_{xx} &=
            \left( \frac{3}{2\pi} \right)^{1/2} A_{1, -1}, \\
        \label{eq:alpha_yy}
        \alpha_{yy} &=
            -\ui \left( \frac{3}{2\pi} \right)^{1/2} A_{1, -1}, \\
        \label{eq:alpha_zz}
        \alpha_{zz} &=
            \left( \frac{3}{4\pi} \right)^{1/2} A_{1, 0},
    \end{align}
\end{subequations}
and $\alpha_{ij} = 0$ for $i \neq j$. Note that the formula for $\alpha_{xx}$ [Eq.~\eqref{eq:alpha_xx}] is only valid for $\Enull{x}$, the formula for $\alpha_{yy}$ is only valid for $\Enull{y}$, and the formula for $\alpha_{zz}$ is only valid for $\Enull{y}$. Hence, each component of the polarizability tensor is determined by one simulation each. This is the reason why Eq.~\eqref{eq:polarizability-from-alm} seems to disagree with Eq.~\eqref{eq:dipole-moment-components}. After determining all components of the dielectric tensor describing the effective medium thin film, the reflectivity of the system can be readily calculated from standard theory~\cite{azzam1987ellipsometry}. In the following, we will denote the surface reflectivity $R_\beta(\omega)$ where the subscript $\beta = \ppol, \spol$ indicates the linear polarization of the incident light. Finally, we note that whereas we can assume $\left| \kp \right| = 0$ for the solution of the Laplace equation, the lateral wave vector enters as
\begin{align*}
    \kp = \frac{\omega}{c} \sin\theta_0 (\cos\phi_0, \sin\phi_0, 0)
\end{align*}
in the reflectance formulas. This is a consequence of the fact that the reflectance in general depends on the angle of incidence~\cite{azzam1987ellipsometry}.



\section{\label{sec:collective-plasmon-modes}The collective plasmon modes of nanoparticle arrays} 

\begin{figure*}[t]
    \centering
    \subfigure{
        \label{fig:dipole-moment-comp-mie}
        \includegraphics{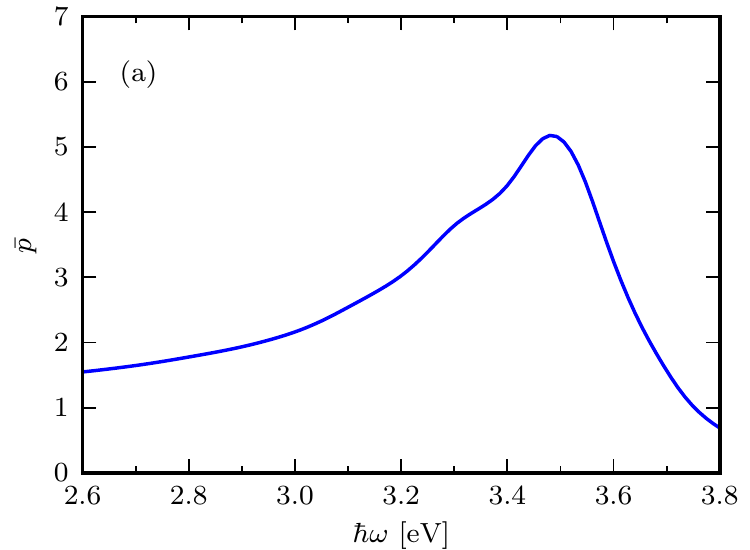}
    }
    \subfigure{
        \label{fig:dipole-moment-comp-monomer}
        \includegraphics{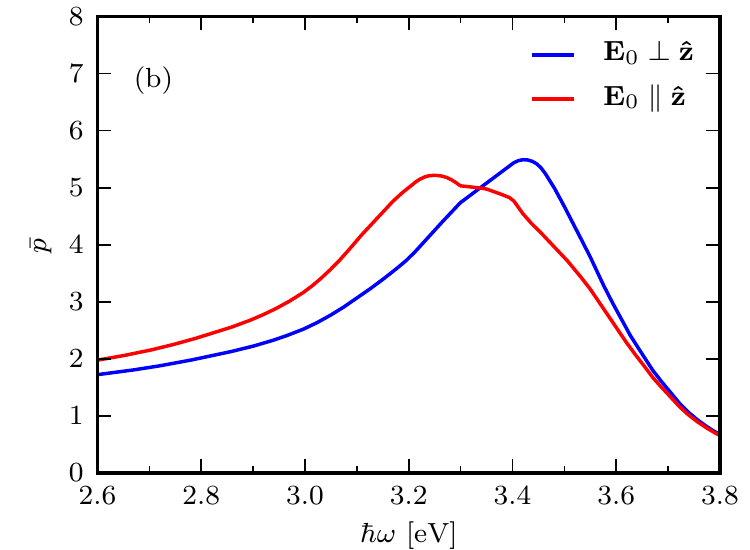}
    }
    \subfigure{
        \label{fig:dipole-moment-comp-dimer}
        \includegraphics{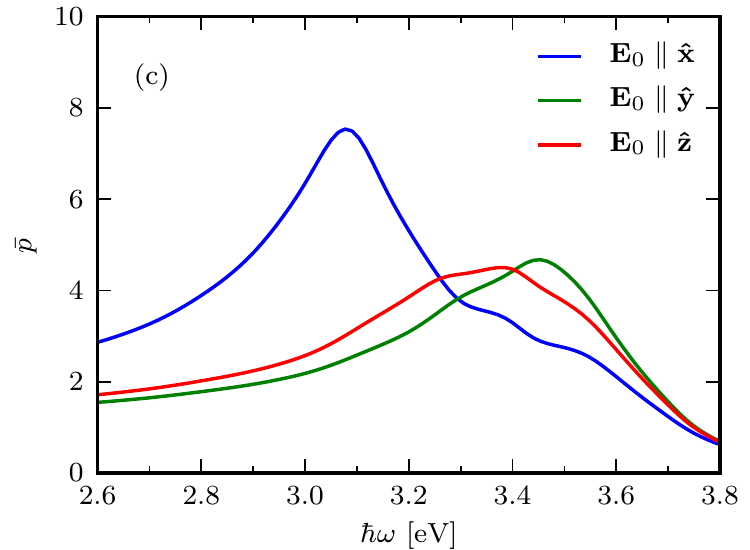}
    }
    \subfigure{
        \label{fig:dipole-moment-comp-lattice}
        \includegraphics{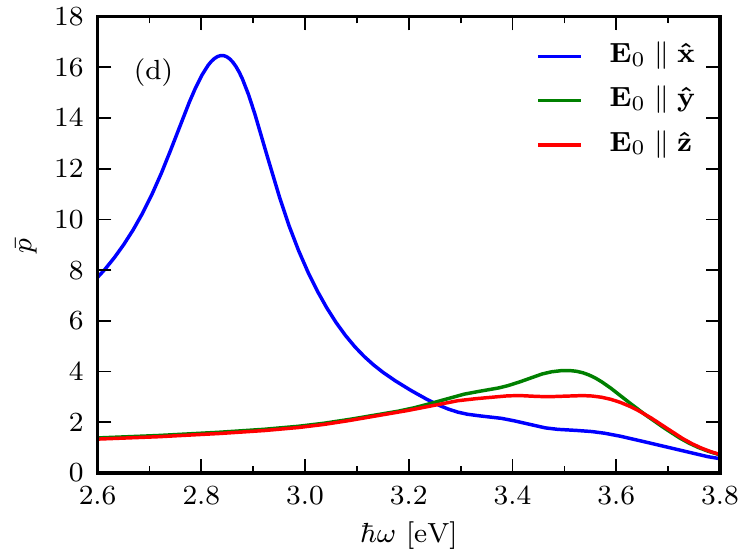}
    }
    \caption{\label{fig:dipole-moment-comp}
        The absolute value of the dimensionless dipole moment of a single nanosphere in the following environments: (a) ``hovering'' in vacuum, (b) supported by an \ce{Al2O3} substrate, (c) in an \ce{Ag} dimer supported by an \ce{Al2O3} substrate and with an interparticle center-center distance of $2.2 a$, (d) in an \ce{Ag} lattice [Figs.~\ref{fig:geometry} and~\ref{fig:geometry_2}] with $b_x = 2.2 a$ and $b_y = 4 b_x$ supported by an \ce{Al2O3} substrate. For comparison, a perfect metal sphere (of radius much smaller than $\lambda$) in vacuum excited by a homogeneous $\tvec{E}_0$ has $\bar{p}(\omega) \equiv 1$. The parameters common to all subfigures were $h = 0.01a$ and $L_{\max} = 30$.
    }
\end{figure*}

In order to estimate how the interactions between the nanoparticles influence their  plasmonic resonances, we compare the dimensionless dipole moment [Eq.~\eqref{eq:p-bar}] of an \ce{Ag} nanoparticle when it is situated in various environments in Fig.~\ref{fig:dipole-moment-comp}. The dipole moment of a single \ce{Ag} nanoparticle (with no substrate present) is shown in Fig.~\ref{fig:dipole-moment-comp-mie} and it exhibits a peak at $\hbar\omega \approx \SI{3.5}{\electronvolt}$. This is the well-known Mie resonance~\cite{Mie:1908mz} in the quasistatic regime.
We note that in obtaining the results of Fig.~\ref{fig:dipole-moment-comp} a cutoff of $L_{\mathrm{max}} = 30$ was used in the calculations. Moreover, the same value of $L_{\mathrm{max}}$ was assumed in obtaining all results that will be presented in this paper.

When the \ce{Ag} monomer is supported ($h = 0.01a$) by a semi-infinite \ce{Al2O3} substrate ($\varepsilon_- \approx 2.76$ at \SI{3}{\electronvolt}, with little dispersion), its dipole moment response is as shown in Fig.~\ref{fig:dipole-moment-comp-monomer}. By comparing Figs.~\ref{fig:dipole-moment-comp-mie} and~\ref{fig:dipole-moment-comp-monomer}, it is observed that the presence of the substrate leads to a redshift of the Mie resonance when the electric field is perpendicular to the substrate (\Enull{z}), but is almost unchanged in the case when  the electric field is parallel to the substrate  ($\mathbf{E}_0 \perp \htvec{z}$). The Mie resonance splits due to the breaking of symmetry caused by the presence of a substrate. Note that in this case, full convergence was not obtained. Nevertheless, the results are shown for easy comparison with the other subfigures of Fig.~\ref{fig:dipole-moment-comp}. For a detailed discussion of this system, see Ref.~\onlinecite{doi:10.1021/jp066539m}.

\begin{figure}[tb]
    \centering
    \includegraphics{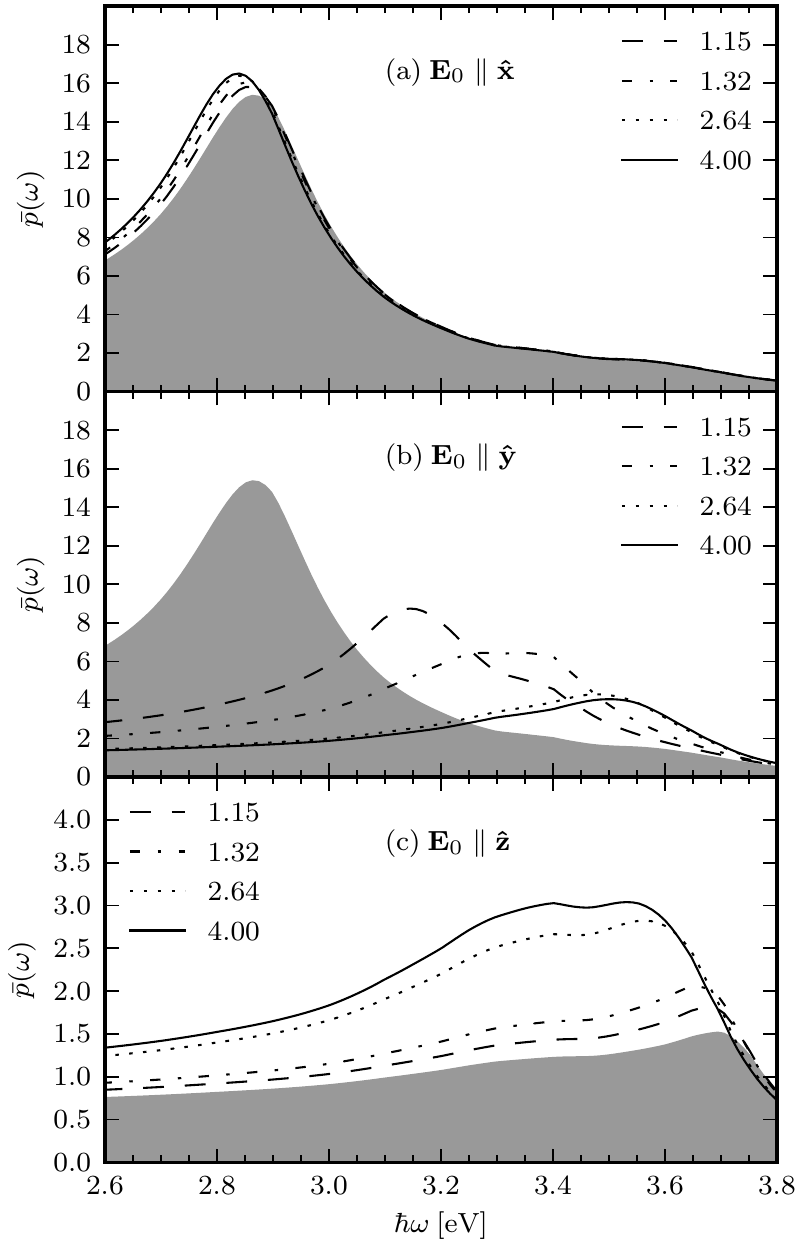}
    \caption{\label{fig:dipole-aniso-scan}
        The absolute value of the dimensionless dipole moment for a sphere in a rectangular lattice, with anisotropy parameters $b_y / b_x = 1.15$ to $b_y / b_x = 4$; $b_x$ was kept constant at $2.2 a$. For reference, the dipole moment of a sphere in a square lattice ($b_x = b_y$) is shown in gray shade. In each plot, the exciting electric field is directed along different axes:
        (a) \Enull{x},
        (b) \Enull{y},
        and
        (c) \Enull{z}.
        Note that the scale of the second axis in (c) is different from the scales used for the equivalent axes of the other two subfigures.
    }
\end{figure}

Next, we consider two supported \ce{Ag} particles placed in  a dimer configuration with the distance between the particle centers being $2.2 a$ and the dimer axis being oriented along the $x$ axis. We also assume that it is ``hovering'' $h = 0.01a$ above an \ce{Al2O3} substrate. We observe that the resonance for $\tvec{E}_0 \perp \htvec{z}$ splits into two resonances, located at different photon energy for \Enull{x} relative to the \Enull{y} case [Fig.~\ref{fig:dipole-moment-comp-dimer}]. This happens because we no longer have rotational symmetry about the $z$ axis. If instead the incident electric field is oriented along the dimer axis (\Enull{x}), one gets a redshift of approximately \SI{0.45}{\electronvolt} (relative to a corresponding isolated particle) and a significant enhancement of the resonance. The particle-substrate interactions are not particularly strong for the \ce{Al2O3} substrate we consider; it only leads to a small redshift and broadening of the resonance when the incident electric field is directed normal to the surface (\Enull{z}). When the incident electric field is directed normal to both the surface of the substrate and the dimer axis, i.e., \Enull{y}, the spectrum looks very similar to the case $\mathbf{E}_0 \perp \htvec{z}$ of an isolated nanoparticle supported by the same substrate [Fig.~\ref{fig:dipole-moment-comp-monomer}].

Finally, we study the case when the neighborhood of the particle becomes a two-dimensional rectangular lattice of (identical) nanoparticles with lattice constants $b_x = 2.2 a$ and $b_y = 4 b_x$, supported by an \ce{Al2O3} substrate ($h = 0.01 a$). In this case, the Mie resonance for \Enull{x} undergoes further redshift and enhancement [Fig.~\ref{fig:dipole-moment-comp-lattice}]. As expected, infinite chains of particles cause stronger interactions between the particles, causing more redshift and resonance enhancement.
When the incident electric field is transverse to the chains and to the surface normal, i.e., \Enull{y}, the interactions between the chains of particles are rather weak and cause no significant redshift. Finally, for $\tvec{E}_0$ normal to the substrate (\Enull{z}), the particle-substrate interactions mainly serve to kill off the resonance. This may be understood by recalling that the induced dipole in the particle and the image dipole in the substrate will have opposite directions and therefore partly cancel the effect of each other.

By comparing the results for the different ratios $b_y / b_x$ shown in Fig.~\ref{fig:dipole-moment-comp}, we see that the local environment can significantly alter the optical response of nanoparticles. In particular, periodic lattices strongly influence the optical response of nanoparticles to an incident electric field. In Sec.~\ref{sec:reflectivity}, we will discuss how this can be observed in a reflectivity experiment.

By studying the dipole moment of a single nanosphere in a rectangular lattice, one seeks a better understanding of how the lattice affects the strength and position of particle resonances and under which conditions the rectangular lattice turns into a collection of non-interacting chains. Hence, in Fig.~\ref{fig:dipole-aniso-scan} we show the (single particle) dipole moment as a function of the anisotropy parameter $b_y / b_x$ of the lattice.
As this parameter increases, the intersphere interactions along the $y$-axis weaken [Fig.~\ref{fig:dipole-aniso-scan}(b)], leading to reduced resonance strength and less redshift of the resonance. We observe that from $b_y / b_x = 2.64$ to $b_y / b_x = 4$, there is little change in both resonance strength and position. Results for intermediate values, i.e., $2.64 < b_y / b_x < 4$, are virtually indistinguishable from those of $b_y / b_x = 4$. From this result, we conclude that for $b_y \geq 4 b_x$, one may ignore interactions between the nanosphere chains altogether (at least for our choice of $b_x$).

When examining Fig.~\ref{fig:dipole-aniso-scan}(a), it is clear that if the electric field is directed along \htvec{x}, the change in $b_y$ does not affect the plasmonic resonances significantly. This is because the interactions between neighbouring particles is strongest along the direction of the incident field. For the case when \Enull{z} [Fig.~\ref{fig:dipole-aniso-scan}(c)], however, the anisotropy affects the plasmonic resonances by increasing the resonance strength. This effect is mainly caused by the reduced density of particles on the substrate, as neighbouring particles cause damping of the resonance. This damping effect can be explained in a ``hand-waving'' fashion by considering the fact that side-by-side parallel dipoles counteract polarization of each other.

Simulations have also been run for lattices where the anisotropy was kept at $b_y / b_x = 4$, but $b_x$ (and thus, $b_y$) was increased. As would be expected, increasing the lattice constants reduces interparticle interactions, and the dipole moment behavior becomes more similar to that of Fig.~\ref{fig:dipole-moment-comp-monomer}. Hence, for lattice constants larger than what has been explored here, one can safely assume that a lattice with $b_y / b_x > 4$ can be approximated as a set of non-interacting linear chains of nanoparticles.

Finally, we have investigated the effects of finite $\left| \kp \right|$. While the effects of a phase difference between neighbouring particles can be significant, we do not include any results in this paper. The reason is that we are primarily interested in optical fields, where $\left| \kp \right| \ll \pi / b_y$. Results of simulations for small, but finite, $\left| \kp \right|$, indicate that the approximation $\left| \kp \right| = 0$ is a good one.


\section{\label{sec:reflectivity}Reflectivity of nanoparticle arrays} 

\begin{figure}[htb]
    \centering
    \subfigure[~$\theta_0 = \ang{0}$]{\label{fig:reflectivity_iso_0deg}
        \includegraphics{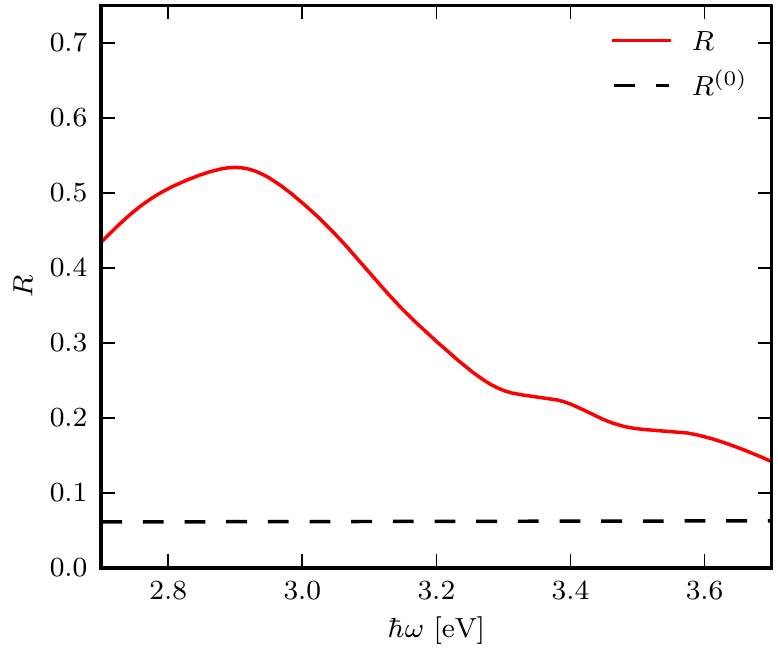}
    }
    \subfigure[~$\theta_0 = \ang{45}$]{\label{fig:reflectivity_iso_45deg}
        \includegraphics{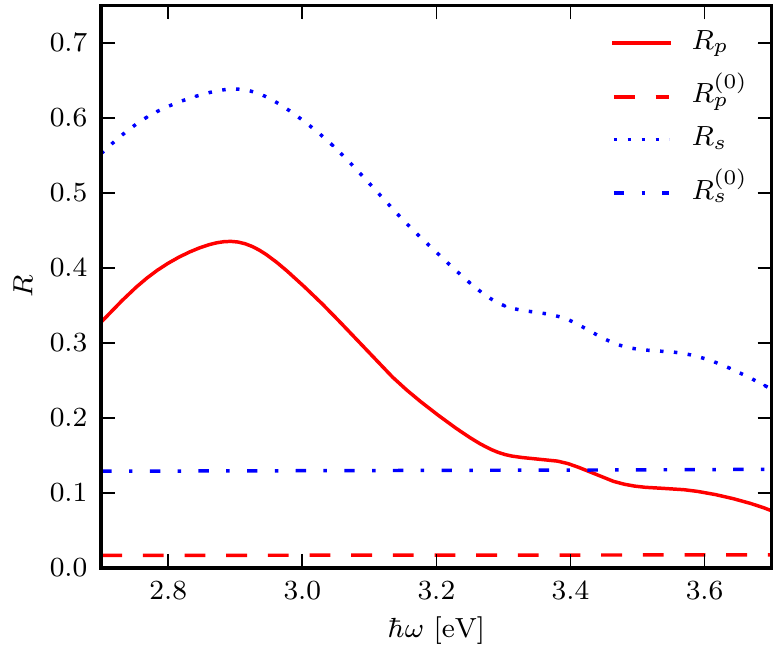}
    }
    \caption{\label{fig:reflectivity_iso}
        Reflectivity of a square lattice of spheres supported by an \ce{Al2O3} substrate for angles of incidence (a) $(\theta_0, \phi_0) = (\ang{0},\ang{0})$ and (b) $(\theta_0,\phi_0) = (\ang{45},\ang{0})$. The lattice parameters were $a = \SI{10}{\nano\meter}$ and $b_x = b_y = 2.2a$. The isotropic nature of the square lattice means that the reflectivity is identical for \ppol-polarized and \spol-polarized light at normal incidence. The dashed lines shows the reflectivity of a substrate with no nanoparticles.
    }
\end{figure}

The reflectivity of an interface is an experimentally accessible quantity for probing the nanoparticle system. Optical methods have the advantage of being non-destructive, and can be used \emph{in situ} in various challenging environments, e.g., vacuum chambers.  For these reasons, we have calculated the surface reflectivity of a square and a rectangular two-dimensional lattice of \ce{Ag} nanoparticles supported by an \ce{Al2O3} substrate.
In these calculations, the incident field is assumed to be a plane wave that is either \ppol{}- or \spol{}-polarized. At normal incidence, we define a \ppol-polarized (\spol-polarized) field by \Enull{x} (\Enull{y}).
The lattice constants were $b_x = b_y = 2.2a$ for the  square lattice and $b_x = 2.2a$, $b_y = 2b_x$ for the  rectangular lattice. In both cases,
$a = \SI{10}{\nano\meter}$ was the radius of the spheres.

\begin{figure}[tb]
    \centering
    \subfigure[~$(\theta_0, \phi_0) = (\ang{0}, \ang{0})$]{\label{fig:reflectivity_aniso_0deg}
        \includegraphics{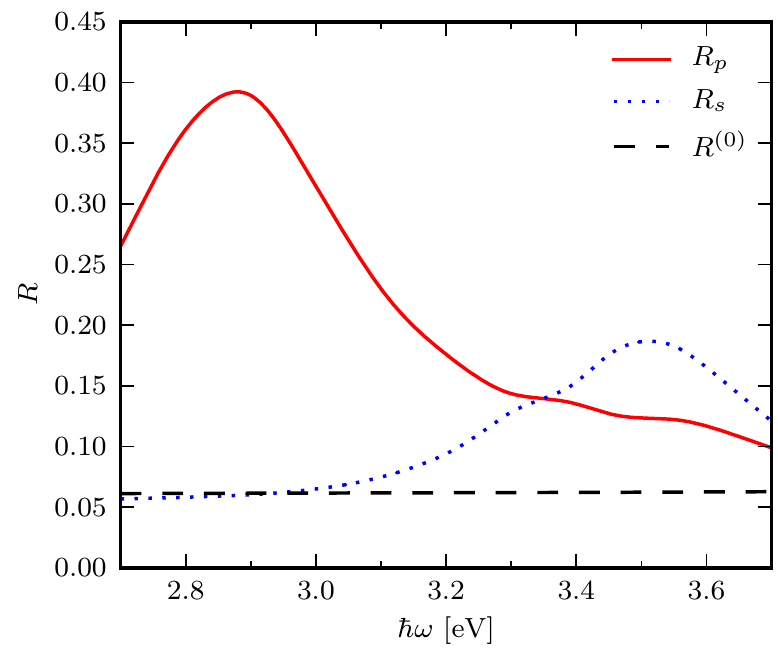}
    }
    \subfigure[~$(\theta_0, \phi_0) = (\ang{45}, \ang{0})$]{\label{fig:reflectivity_aniso_45deg}
    \includegraphics{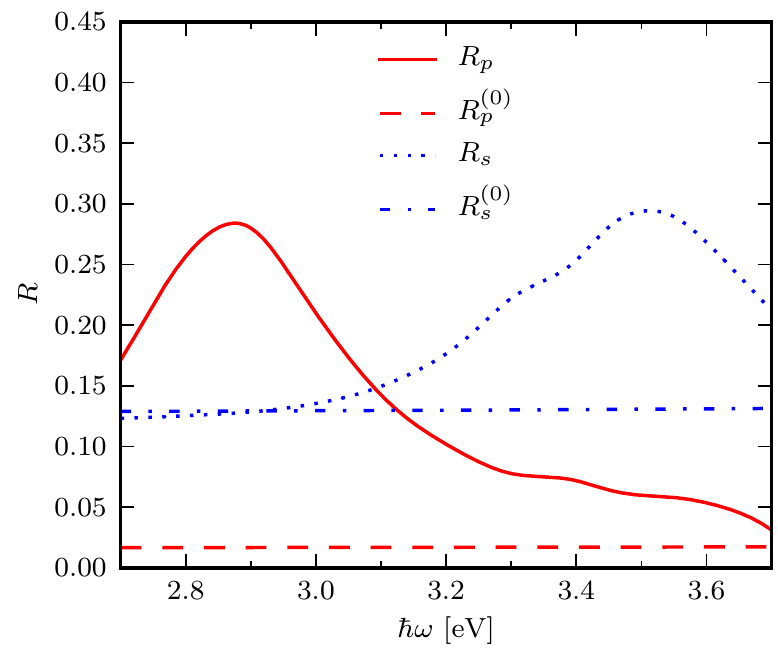}
    }
    \caption{\label{fig:reflectivity_aniso}
        Reflectivity of a rectangular lattice of spheres on top of an \ce{Al2O3} substrate. The lattice parameters were $a = \SI{10}{\nano\meter}$, $b_x = 2.2a$, and $b_y = 2b_x$. The dashed lines shows the reflectivity of a substrate with no nanoparticles. (a) At normal incidence [$(\theta_0,\phi_0) = (\ang{0},\ang{0})$] we define \ppol-polarized light to be polarized along the $x$-axis, and (b) $(\theta_0, \phi_0) = (\ang{45}, \ang{0})$.
    }
\end{figure}

\begin{figure}[tb]
    \centering
    \subfigure[~$\theta_0 = \ang{0}$]{\label{fig:thin-film-reflectivity-0deg}
        \includegraphics{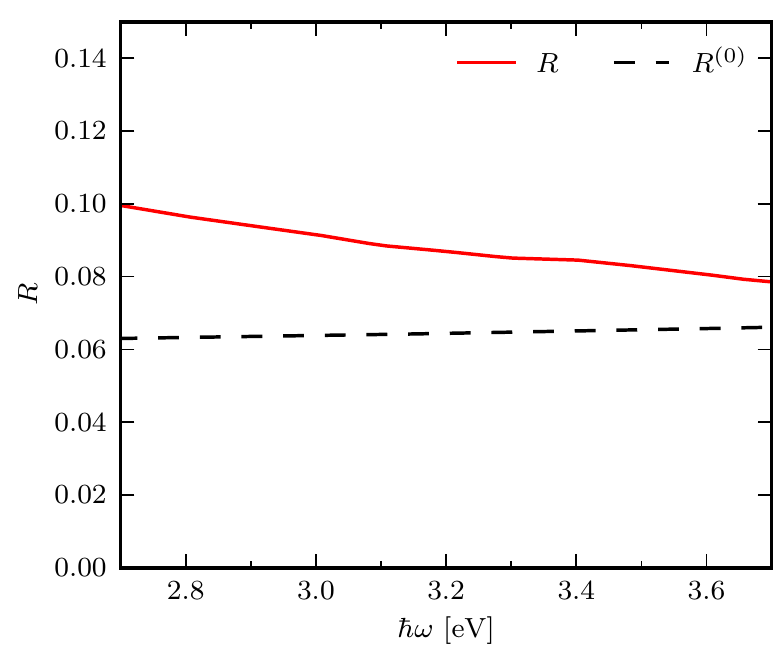}
    }
    \subfigure[~$\theta_0 = \ang{45}$]{\label{fig:thin-film-reflectivity-45deg}
        \includegraphics{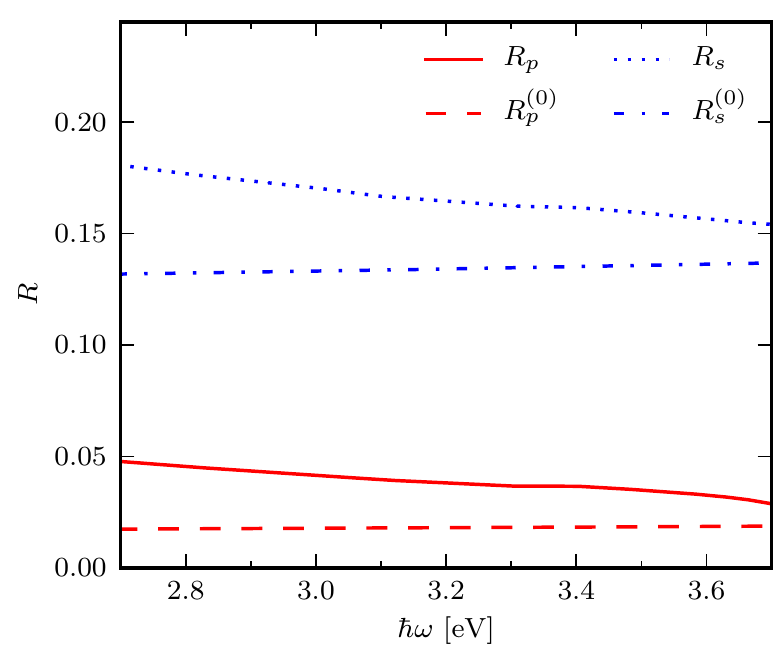}
    }
    \caption{\label{fig:thin-film-reflectivity}
      The reflectivity of an isotropic and homogeneous \ce{Ag} film at angles of incidence (a) $\theta_0 = \ang{0}$ and (b) $\theta_0 = \ang{45}$ (as the film is isotropic, $\phi_0$ is irrelevant). The film thickness was assumed to be \SI{4.3}{\nano\meter}, which is equivalent to the mass thickness of the rectangular lattice discussed in Fig.~\ref{fig:reflectivity_aniso}.
    }
\end{figure}

First we address normal incidence. Figures~\ref{fig:reflectivity_iso_0deg} and~\ref{fig:reflectivity_aniso_0deg} present the calculated reflectivity (Sec.~\ref{sub:reflectivity}) for the square and rectangular lattices, respectively, at angles of incidence $(\theta_0,\phi_0) = (\ang{0},\ang{0})$.
The red solid curve in Fig.~\ref{fig:reflectivity_iso_0deg} depicts the reflectivity from the square lattice at normal incidence, whereas the dashed line (labeled $R^{(0)}$) shows the reflectivity from a flat, clean \ce{Al2O3} surface. From this figure it is observed that there is no difference (at normal incidence) between \ppol- and \spol-polarized reflected light, as is to be expected due to the symmetry of the lattice.
Moreover, the presence of the metallic particles cause the reflectivity to increase relative to the reflectivity of a clean dielectric substrate. This increase is especially pronounced near the redshifted Mie resonance.

For the rectangular lattice at normal incidence, the reflectivity of \ppol{}-polarized light looks rather different from the reflectivity of \spol{}-polarized light~[Fig.~\ref{fig:reflectivity_aniso_0deg}], caused by $b_x \neq b_y$. Since $b_y = 2 b_x$ was assumed, the particles will interact more strongly along the $x$-direction than along the $y$-direction.
Based on what was found in Fig.~\ref{fig:dipole-aniso-scan}, the system can be thought of as a set of weakly interacting linear chains oriented along the $x$-axis. Along the chain, the interactions are strong (\ppol-polarized light has a component along $\htvec{x}$), whereas between the chains (\spol-polarized light is polarized along $\htvec{y}$) the interactions are much weaker.
This can be seen directly in the reflectivity curves: For \ppol{} polarization, a large redshift relative to the isolated, single particle Mie resonance is observed due (mainly) to interparticle interactions. However, for \spol{} polarization there is essentially no redshift, meaning that the resonance occurs close to the single particle Mie resonance. Note that the interparticle interactions along the chain ($\htvec{x}$) in \ppol{} polarization produce a much stronger resonance, and hence higher reflectivity, than the interchain interactions in \spol{} polarization.

We now turn to non-normal incidence, $(\theta_0, \phi_0) = (\ang{45}, \ang{0})$, for which results are presented in Figs.~\ref{fig:reflectivity_iso_45deg} and~\ref{fig:reflectivity_aniso_45deg}. In this case, the two polarization states give rise to different reflectivity curves even for the square lattice. From  Fig.~\ref{fig:reflectivity_iso_45deg}, it is observed that the reflectivity for \spol{}-polarized light is systematically higher than the reflectivity for \ppol{}-polarized light.
Moreover, the difference between the two curves is almost independent of energy, and the relative change in the reflectivity as a function of energy follows closely that of normal incidence. This behavior is due to the difference in reflectivity at non-normal incidence for the two linear polarizations, as indicated by the dashed lines in Figs.~\ref{fig:reflectivity_iso_45deg} and~\ref{fig:reflectivity_aniso_45deg}.

For the same angles of incidence [$(\theta_0, \phi_0) = (\ang{45}, \ang{0})$], Fig.~\ref{fig:reflectivity_aniso_45deg} shows that the reflectivity from the rectangular lattice is both quantitatively and qualitatively different from that of the square lattice. As the interactions between the particles are now significantly stronger for electric fields polarized along the $x$ axis, the \ppol{}-polarized reflectivity still has a significantly redshifted peak.
For \spol{}-polarized light, however, the redshift is negligible. This means that this surface acts as a ``spectral polarizer'': It reflects predominantly \ppol{} polarized light at around 2.9 eV, whereas at around 3.5 eV, \spol{}-polarized light dominates. The reflectivity can be increased by simply using a greater ``mass thickness'' (the equivalent thickness of a homogeneous \ce{Ag} thin film) of \ce{Ag}, i.e., by using larger nanoparticles. However, the results will be affected by retardation effects if the nanoparticle radius increases too much.

For comparison, we present in Fig.~\ref{fig:thin-film-reflectivity} the reflectivity of a continuous, isotropic, and homogeneous \ce{Ag} film of thickness \SI{4.3}{\nano\meter}. This is the equivalent mass thickness of the rectangular lattice with $b_y = 2 b_x$. By comparing the results of Figs.~\ref{fig:reflectivity_aniso_45deg} and~\ref{fig:thin-film-reflectivity} it is readily observed that the corresponding reflectivity curves are rather different. Thus, it is feasible to separate the case of a continuous thin film from that of isolated metallic island films through reflectance measurements.
In fact, the reflectivity of the thin continuous film is much closer to the reflectivity of the clean substrate, where few features of interest can be observed.


\section{\label{sec:Conclusion}Conclusion} 

In this paper, we have investigated the collective excitations in square and rectangular two-dimensional lattices of \ce{Ag} nanoparticles supported by a dielectric (\ce{Al2O3}) substrate. In particular, we have established that for lattices where the lattice constant is greater than approximately $4$ sphere diameters ($b_y / b_x > 4$), the system can be approximated as a collection of independent, non-interacting linear chains. It has also been demonstrated that if the incident field is polarized along the shortest lattice vector, the anisotropic nature of the lattice leads to collective resonances that are significantly redshifted relative to the single particle Mie resonance.

We have also presented results showing the reflectivity of surfaces patterned by such nanoparticle lattices. If the lattice is rectangular, the surfaces exhibit the interesting property that the reflected light possesses different colors in the two linear polarizations (\ppol{} and \spol{}).
It is reasonable to assume that similar behavior will be apparent in transmission. For technological applications, the advantage of the rectangular lattice configuration is that no control is necessary over the nanoparticle orientation (cf. anisotropic particles on a square lattice), and that resonance positions can be tuned via the lattice constants.

Since rectangular lattices of nanoparticles leave much room for tunability of the plasmonic and polarization characteristics of an interface, we believe further studies are in order to gain more insight. Tunability can further be extended by employing other materials; nanoshells, which allow for tunability through the core and shell radii; or through use of truncated spheres or otherwise anisotropic particles. For spherical nanoshells, the theory sketched in this paper could be applied with relatively straightforward modifications. The effects of randomness on the properties of such lattices is also a topic of interest~\cite{doi:10.1021/la300198r}, relevant to experimental conditions in which some element of randomness is inevitable. The approach here is not very computationally demanding, and should thus be suited for exploration of random effects.


\begin{acknowledgements} 

The authors would like to acknowledge the Norwegian University of Science and Technology for the allocation of computer time.

P.A.L. would like to thank Dr.~Ingar S. Nerb\o{}, Lars Martin Sandvik Aas, Tor Nordam, and Dr.~Jamie Cole for interesting discussions. Moreover, I.S. acknowledges fruitful discussions with Dr. R\'emi Lazzari.

We used the freely available computer library \textsc{SHTOOLS} authored by Dr.~Mark Wieczorek to evaluate the associated Legendre functions.

\end{acknowledgements} 

\end{document}